# Regional airports in Greece, their characteristics and their importance for the local economic development

**Serafeim Polyzos and Dimitrios Tsiotas**

University of Thessaly, Polytechnic School,
Department of Planning and Regional Development,
Pedion Areos, Volos,
tel. 24210 74446
spolyzos@uth.gr; tsiotas@uth.gr

**Abstract**

Technological developments worldwide are contributing to the improvement of transport infrastructures and they are helping to reduce the overall transport costs. At the same time, such developments along with the reduction in transport costs are affecting the spatial interdependence between the regions and countries, a fact inducing significant effects on their economies and, in general, on their growth-rates. A specific class of transport infrastructures contributing significantly to overcoming the spatial constraints is the air-transport infrastructures. Nowadays, the importance of air-transport infrastructures in the economic development is determinative, especially for the geographically isolated regions, such as for the island regions of Greece. Within this context, this paper studies the Greek airports and particularly the evolution of their overall transportation imprint, their geographical distribution, and the volume of the transport activity of each airport. Also, it discusses, in a broad context, the seasonality of the Greek airport activity, the importance of the airports for the local and regional development, and it formulates general conclusions.

# Τα περιφερειακά αεροδρόμια στην Ελλάδα, τα χαρακτηριστικά τους και η σημασία τους για την τοπική οικονομική ανάπτυξη

## Σεραφείμ Πολύζος και Δημήτριος Τσιώτας

Πανεπιστήμιο Θεσσαλίας, Πολυτεχνική Σχολή,
Τμήμα Μηχανικών Χωροταξίας, Πολεοδομίας και Περιφερειακής Ανάπτυξης,
Πεδίο Άρεως, Βόλος,
τηλ. 24210 74446
spolyzos@uth.gr; tsiotas@uth.gr

**Περίληψη**

Οι τεχνολογικές εξελίξεις που πραγματοποιούνται σε όλες τις χώρες επιδρούν στη βελτίωση των μεταφορικών υποδομών, συμβάλλοντας στη μείωση του γενικευμένου μεταφορικού κόστους. Ταυτόχρονα, οι εξελίξεις αυτές και η μείωση του μεταφορικού κόστους συμβάλλουν στην ένταση της χωρικής αλληλεξάρτησης μεταξύ των περιφερειών και των χωρών, με σημαντικές επιδράσεις στις οικονομίες τους και γενικότερα στους ρυθμούς ανάπτυξής τους. Μια ιδιαίτερη κατηγορία μεταφορικών υποδομών με σημαντική συμβολή στην υπέρβαση των γεωγραφικών «εμποδίων» είναι οι αεροπορικές υποδομές. Στη σύγχρονη εποχή η σημασία των αεροπορικών υποδομών στην οικονομική ανάπτυξη μπορεί να χαρακτηριστεί ως καθοριστική, ειδικότερα για τις γεωγραφικά απομονωμένες περιοχές, όπως είναι οι νησιωτικές περιοχές της Ελλάδας. Με τα αεροδρόμια της Ελλάδας και συγκεκριμένα την εξέλιξη του συνολικού μεταφορικού τους έργου, τη γεωγραφική τους κατανομή και το μέγεθος της μεταφορικής δραστηριότητας κάθε αεροδρομίου ασχολείται το άρθρο αυτό. Επίσης, στο άρθρο αναλύονται σε ένα ευρύ πλαίσιο η εποχικότητα στη δραστηριότητα των αεροδρομίων της Ελλάδας, η σημασία των αεροδρομίων για την τοπική και περιφερειακή ανάπτυξη και διατυπώνονται γενικά συμπεράσματα.

**Λέξεις κλειδιά:** περιφερειακή οικονομική, ανισότητες, κύρτωση, συντελεστής Gini

## 1. Εισαγωγή

Οι ανθρώπινες δραστηριότητες πραγματοποιούνται σε διαφορετικά σημεία του γεωγραφικού χώρου. Αυτό δημιουργεί την ανάγκη των μετακινήσεων του πληθυσμού από περιοχή σε περιοχή και των μεταφορών των αγαθών από τις περιοχές ή τα σημεία παραγωγής προς τις περιοχές ή τα σημεία κατανάλωσης. Συνέπεια των παραπάνω είναι η ανάπτυξη και λειτουργία των μεταφορικών υποδομών, οι οποίες εξυπηρετούν τις εν λόγω μετακινήσεις και μεταφορές. Γενικότερα, οι μεταφορικές υποδομές και τα μεταφορικά μέσα αποτελούν βασικούς παράγοντες που δια μέσου της μεταφοράς ανθρώπων, αγαθών, πληροφοριών, γνώσεων και τεχνολογίας επιτρέπουν την επικοινωνία και την διατοπική σύνδεση (Πολύζος 2019a, Πολύζος 2019c). Με την πάροδο του χρόνου και υπό την επίδραση των τεχνολογικών εξελίξεων οι μεταφορικές υποδομές και τα μεταφορικά μέσα βελτιώνονται, με αποτέλεσμα τη συνεχή μείωση των χρονο-αποστάσεων και του γενικευμένου μεταφορικού κόστους, ώστε να μπορεί να χρησιμοποιηθεί η έκφραση: «..*η γη συνεχώς μικραίνει..*». Η «συρρίκνωση» της γεωγραφικής απόστασης, την οποία επιτυγχάνουν οι μεταφορικές υποδομές, έχει ως συνέπεια τη μεταβολή, κατά κανόνα αύξηση, του βαθμού ή της έντασης της αλληλεπίδρασης των δραστηριοτήτων των περιφερειών, που εκφράζονται με τις διαπεριφερειακές άυλες ή υλικές «ροές» (Tsiotas and Polyzos, 2018). Αποτέλεσμα των «χωρικών ιδιοτήτων» των μεταφορικών υποδομών είναι οι σημαντικές επιδράσεις τους στη οικονομική αλληλεξάρτηση των περιφερειών και κατ' επέκταση στην οικονομική τους ανάπτυξη (Polyzos, 2009; Πολύζος, 2015; Πολύζος 2019b).

Μια ιδιαίτερη κατηγορία μεταφορικών υποδομών είναι τα αεροδρόμια. Στη σύγχρονη εποχή τα αεροδρόμια κατέχουν ηγετικό ρόλο στις εξελίξεις που χαρακτηρίζουν την παγκοσμιοποίηση, ενώ η σημασία των αεροπορικών μεταφορών για την οικονομική ανάπτυξη των χωρών και των περιφερειών συνεχώς αυξάνεται (Andrew and Bailey, 1996; Green, 2007; Mukkala and Tervo, 2013). Οι αεροπορικές μεταφορές αναπτύσσονται με ραγδαίους ρυθμούς σε όλες τις χώρες και λόγω της ταχύτητας που προσφέρουν, διευκολύνουν σημαντικά τη μετακίνηση επιβατών και αγαθών σε προορισμούς υπερτοπικής εμβέλειας (Savage, 2013; Zak and Getzner, 2014; Tsiotas and Polyzos, 2015; Tsiotas et al., 2019a).

Παρά το γεγονός ότι το κόστος των αεροπορικών μεταφορών είναι μεγαλύτερο σε σχέση με τα άλλα μεταφορικά μέσα, οι αεροπορικές μεταφορές είναι γενικά ασφαλέστερες και ταχύτερες. Σε αρκετές περιπτώσεις αποτελούν τη μοναδική ή τη βασική επιλογή για τη σύνδεση απομονωμένων αγροτικών περιοχών και νησιών με αστικές περιοχές ή για τη σύνδεση αμοιβαία απομακρυσμένων τοποθεσιών (Savage, 2013). Οι αεροπορικές μεταφορές έχουν ως βασικό πλεονέκτημα τη μεγάλη ταχύτητα μεταφοράς ανθρώπων ή εμπορευμάτων, ενώ τα μειονεκτήματά τους αφορούν στο μεγάλο κόστος μεταφοράς, στην αδυναμία μεταφοράς ογκωδών ή εμπορευμάτων μεγάλου βάρους και τη μικρή ευελιξία τους (Andrew and Bailey, 1996; Mukkala and Tervo, 2013).

Οι τεχνολογικές εξελίξεις οδήγησαν μετά τη δεκαετία του '60 σε μεγάλη αύξηση των αεροπορικών μεταφορών σε σχέση με τις θαλάσσιες, σιδηροδρομικές ή οδικές μεταφορές. Η αυξητική τάση συνεχίστηκε και τις επόμενες δεκαετίες, με τριπλασιασμό της αεροπορικής μεταφοράς επιβατών τη δεκαετία 1970-1979, κάτι που οφείλονταν στις εξελίξεις της τεχνολογίας, αλλά και στις πολιτικές που εφαρμόσθηκαν για τις μεταφορές στο νέο διεθνοποιημένο περιβάλλον. Την περίοδο 1980-89 η αύξηση στις αεροπορικές μεταφορές ήταν περίπου ίση με 83%, ενώ έκτοτε η επιβατική κίνηση αυξάνεται διαρκώς με ετήσιο μέσο όρο περίπου ίσο με 6%-7%. Ενδεικτικά αναφέρεται ότι, το 2015 η παγκόσμια αεροπορική διακίνηση επιβατών ήταν 6,4% υψηλότερη σε σχέση με το 2014, ενώ οι μετακινηθέντες αεροπορικώς επιβάτες υπερέβησαν τα 7 δισεκατομμύρια, έναντι 6,8 δισεκατομμυρίων το 2014 (EC, 2017).

Πρόκειται για τον υψηλότερο ετήσιο ρυθμό ανάπτυξης από το 2010, το έτος που ακολούθησε την παγκόσμια ύφεση του 2008-09. Την περίοδο αυτή οι συνολικές κινήσεις αεροσκαφών αυξήθηκαν ετησίως κατά 2% και ανήλθαν σε 88 εκατομμύρια το 2015 (EC, 2017). Αξίζει να σημειωθεί ότι, σύμφωνα με στοιχεία της International Air Transport Association (IATA), το 2018 το μέσο κόστος των αεροπορικών μεταφορών ήταν το μισό του προ εικοσαετίας αντίστοιχου κόστους. Επίσης, από το 2010 το αποτύπωμα άνθρακα (carbon footprint) ανά επιβάτη μειώνεται κατά 2,8% ετησίως, οι συνολικές εμπορευματικές μεταφορές παγκοσμίως το 2018 ανήλθαν σε 64 εκατομμύρια τόνους, που αντιστοιχεί στο 1/3 της αξίας του παγκόσμιου εμπορίου, ενώ δημιουργήθηκαν 65 εκατομμύρια θέσεις εργασίας και παρήχθη ΑΕΠ ίσο με 2,7 τρισεκατομμύρια δολάρια (IATA, 2019).

Σύμφωνα με τις προβλέψεις, έως το 2040 η ζήτηση για αεροπορικές μεταφορές αναμένεται να διπλασιαστεί (IATA, 2019). Η αύξηση αυτή θα απαιτήσει ανάλογες μεταβολές και βελτιώσεις στις αεροπορικές υποδομές και τα μέσα μεταφοράς. Η κάλυψη της ζήτησης για αεροπορική συνδεσιμότητα εξαρτάται από τη διαθεσιμότητα των απαραίτητων υποδομών, ενώ θα πρέπει να ληφθούν σχετικές αποφάσεις για τη διαχείριση της εναέριας κυκλοφορίας, τα διαχειριστικά προβλήματα εντός των αεροδρομίων, την ανάπτυξη των υποστηρικτικών υπηρεσιών στις αεροπορικές μεταφορές, το θεσμικό και οργανωτικό πλαίσιο της αεροπορικής αγοράς κ.λπ. (IATA, 2019). Αναφορικά με τις χώρες της Ευρωπαϊκής Ένωσης (ΕΕ-28), σύμφωνα με τα στοιχεία της EUROSTAT, το αεροδρόμιο Heathrow του Λονδίνου ήταν το πιο πολυσύχναστο ως προς τους αριθμούς επιβατών που διακινήθηκαν το 2016, με 76 εκατομμύρια αφίξεις και αναχωρήσεις επιβατών, ενώ παραμένει σταθερά ο πλέον πολυσύχναστος αερολιμένας στην ΕΕ από την έναρξη των χρονολογικών σειρών το 1993. Ακολούθησαν τα αεροδρόμια Charles de Gaulle του Παρισιού με 66 εκατομμύρια διακινήσεις, Schiphol του Άμστερνταμ με 64 εκατομμύρια διακινήσεις και το αεροδρόμιο της Φρανκφούρτης με 61 εκατομμύρια διακινήσεις (EUROSTAT, 2018). Σημειώνεται ότι το μεγαλύτερο ποσοστό των διακινηθέντων μέσω των παραπάνω αεροδρομίων επιβατών αφορούσαν διεθνείς πτήσεις.

Για την ΕΕ-28 κατά μέσο όρο μεταφέρθηκαν αεροπορικώς 1,9 επιβάτες ανά κάτοικο το 2016. Λαμβάνοντας υπόψη το μέγεθος του πληθυσμού κάθε χώρας της ΕΕ-28, η σημασία των αεροπορικών μετακινήσεων ήταν ιδιαίτερα υψηλή για τα νησιά της Μάλτας και της Κύπρου, αφού διακινήθηκαν 11,0 και 10,5 επιβάτες αντίστοιχα ανά κάτοικο το 2016, την Ισλανδία με 20,3 επιβάτες ανά κάτοικο και τη Νορβηγία με 7,2 επιβάτες ανά κάτοικο. Τα χαμηλότερα ποσοστά καταγράφηκαν για δέκα κράτη μέλη από την Ανατολική Ευρώπη και τη Βαλτική, σε καθένα από τα οποία καταγράφηκε μέσος όρος αεροπορικών μετακινήσεων χαμηλότερος από 2,0 επιβάτες ανά κάτοικο το 2016 (EUROSTAT, 2018).

Τα προαναφερθέντα στοιχεία αναδεικνύουν τη σημαντικότητα των αεροπορικών μεταφορών για την οικονομική ανάπτυξη όλων των χωρών. Αναφορικά με την Ελλάδα, ο τομέας των αεροπορικών μεταφορών είναι ιδιαίτερα σημαντικός για την οικονομία και την ανάπτυξη πολλών περιοχών της. Η ιδιαίτερη γεωμορφολογία της Ελλάδας σε σχέση με τις άλλες ευρωπαϊκές χώρες, η οποία περιλαμβάνει ορεινούς όγκους στην ηπειρωτική χώρα και μεγάλο αριθμό μικρών νησιών στο θαλάσσιο χώρο της αποτέλεσε τη βασική αιτία που συνέβαλε στην ανάπτυξη ενός σχετικά προβληματικού συστήματος χερσαίων και θαλάσσιων μεταφορών. Από την άλλη πλευρά, τα παραπάνω χαρακτηριστικά της χώρας οδήγησαν στην ανάπτυξη πολλών περιφερειακών αεροδρομίων (Σκάγιαννης, 2008).

Με δεδομένα τη μεγάλη χωρική διασπορά των νησιών της χώρας και το γεγονός ότι οι βασικές οικονομικές δραστηριότητες των νησιών σχετίζονται με τον τουρισμό, γίνεται αντιληπτό πόσο απαραίτητες είναι οι υποδομές των αεροδρομίων στην Ελλάδα για τη διατήρηση της εδαφικής, κοινωνικής και οικονομικής συνοχής της. Η σημασία των αεροδρομίων για την τοπική ανάπτυξη των γεωγραφικά απομονωμένων περιοχών είναι αδιαμφισβήτητα μεγάλη, καθώς μέσω αυτών επιτυγχάνεται η προσβασιμότητα τους και αίρεται η γεωγραφική τους απομόνωση.

Οι οικονομίες των νησιωτικών περιοχών της Ελλάδας έχουν μεγάλη εξάρτηση από τον τουρισμό (Πολύζος 2019a). Επιπλέον, ο τουρισμός των νησιωτικών περιοχών της Ελλάδας είναι στο μεγαλύτερο μέρος του διεθνής, με συνέπεια να στηρίζεται σε μεγάλο βαθμό στις επιβατικές αερομεταφορές, οι οποίες δεν μπορούν να πραγματοποιηθούν χωρίς σύγχρονα αεροδρόμια. Λόγω της ιδιαίτερης γεωμορφολογίας οι αερομεταφορές κατέχουν ιδιαίτερη θέση στο σύστημα μεταφορών της χώρας, η σημαντικότητα των οποίων είναι καθοριστική τη θερινή περίοδο, αφού περίπου το 70% της τουριστικής κίνησης πραγματοποιείται αεροπορικώς (ΕΛΣΤΑΤ, 2017).

Η χωρική κατανομή των περιφερειακών αεροδρομίων στην Ελλάδα, η εξέλιξη του μεταφορικού έργου που πραγματοποιείται αεροπορικώς, τα γενικότερα χαρακτηριστικά τους και η σημασία τους για την τοπική ανάπτυξη θα αναλυθούν στο άρθρο αυτό. Συγκριμένα, στην επόμενη ενότητα θα αναλυθούν οι μεταβολές στο επιβατικό και εμπορευματικό μεταφορικό έργο των αεροδρομίων στην Ελλάδα. Στη συνέχεια θα γίνει η απεικόνιση της χωρικής κατανομής και της μεταφορικής δραστηριότητας των εν λόγω αεροδρομίων και θα αναλυθεί το πρόβλημα της εποχικότητας των αερομεταφορών. Τέλος θα γίνει αναφορά στη σχέση αεροδρομίων και τοπικής ανάπτυξης και στην τελευταία ενότητα θα διατυπωθούν τα τελικά συμπεράσματα που προκύπτουν από την ανάλυση που προηγήθηκε.

## 2. Η εξέλιξη του αεροπορικού μεταφορικού έργου στην Ελλάδα

Οι τεχνολογικές καινοτομίες προσδίδουν στις αεροπορικές μεταφορές πλεονεκτήματα σε σχέση με τους άλλους τύπους μεταφορών, όπως είναι η πραγματοποίηση δρομολογίων ημέρα και νύχτα, μεγαλύτερη ασφάλεια στη μεταφορά, υπερατλαντικές μεταφορές κ.α. Τα πλεονεκτήματα αυτά βοηθούν στη μείωση του κόστους μεταφοράς σε μεγάλες αποστάσεις συγκριτικά με τα υπόλοιπα μεταφορικά μέσα (Wei and Yanji, 2006; Harrigan, 2010). Με δεδομένο ότι οι αερομεταφορές προκάλεσαν «επανάσταση» στον τομέα των μεταφορών, αφού οι μεγάλες μειώσεις που επέφεραν στις αποστάσεις λόγω της μεγάλης ταχύτητας που αναπτύσσουν τα αεροπλάνα και της δυνατότητας που έχουν για υπέρβαση των περιορισμών του εδαφικού ανάγλυφου, η Ελλάδα είναι από τις χώρες που μπορεί να επιλύσει πολλά προβλήματα με την ανάπτυξη αυτού του τύπου μεταφορών.

Θα αναλυθεί στη συνέχεια το μεταφορικό έργο που πραγματοποιήθηκε αεροπορικώς στην Ελλάδα την 23-ετία 1996-2018. Για τη γραφική απεικόνιση της αεροπορικής κίνησης που εμφάνισαν τα αεροδρόμια της Ελλάδας την περίοδο αυτή, έχουν γίνει τα Διαγράμματα 1, 2 και 3 με χρήση στατιστικών στοιχείων της Υπηρεσίας Πολιτικής Αεροπορίας (ΥΠΑ, 2019). Στα διαγράμματα αυτά απεικονίζονται οι αεροπορικές πτήσεις που πραγματοποιήθηκαν, οι μετακινηθέντες επιβάτες και τα μεταφερθέντα εμπορεύματα.

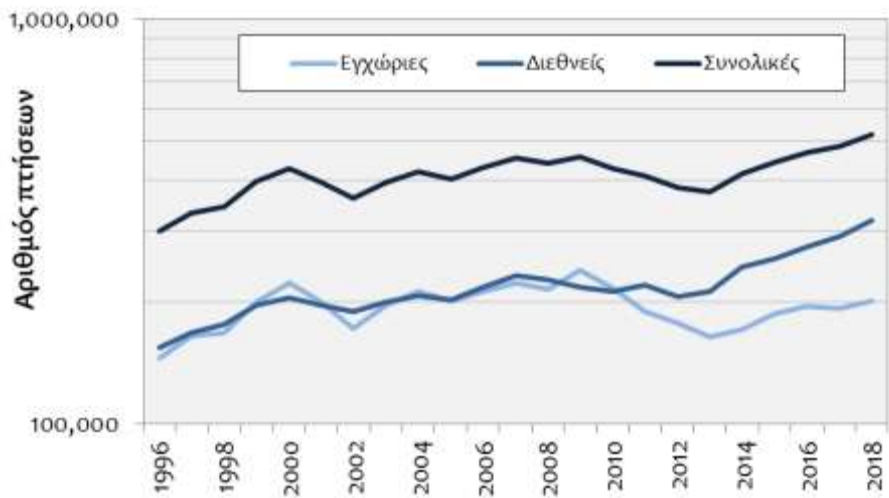

**Διάγραμμα 1:** Οι πτήσεις που πραγματοποιήθηκαν στα αεροδρόμια της Ελλάδας την περίοδο 1996-2018 (οι τιμές στον κάθετο άξονα εμφανίζονται σε λογαριθμική κλίμακα $x \cdot 10^5$, όπου $x=1,2,\ldots,10$ είναι οι δευτερεύουσες γραμμές πλέγματος).

Στο Διάγραμμα 1 παρατηρούμε ότι οι εγχώριες πτήσεις ήταν περίπου ίσες με τις διεθνείς κατά την περίοδο 1996-2009. Με την έναρξη της οικονομικής κρίσης το έτος 2009 υπήρξε μια μείωση των εγχώριων πτήσεων έως το 2013 και στη συνέχεια μια αύξηση. Αντίθετα, οι διεθνείς πτήσεις ακολούθησαν μια συνεχή τάση ανόδου, γεγονός που δείχνει ότι δεν επηρεάστηκαν από την οικονομική κρίση, ενώ από το 2013 και μετά η άνοδος είναι σημαντική. Ο συνολικός αριθμός των πτήσεων που πραγματοποιήθηκαν στις χρονικές περιόδους 1996-2000 και 2013-2018 εμφάνισε ανοδική τάση, ενώ στις άλλες χρονικές μια σταθερή ή καθοδική πορεία. Αναφορικά με τον αριθμό των επιβατών που μετακινήθηκαν κατά την περίοδο 1996-2018, αυτός εμφανίζει ανάλογη μεταβολή με τον αριθμό των αεροπορικών πτήσεων. Ο αριθμός των επιβατών στις πτήσεις του εσωτερικού έχει μια σταθερή εξέλιξη έως το 2009, μια μείωση την περίοδο 2009 έως 2013 και μια αύξηση στη συνέχεια. Ο αριθμός των επιβατών στις πτήσεις του εξωτερικού εμφάνισε μια ανοδική τάση σε όλη τη χρονική περίοδο, ενώ η τάση αυτή είναι ιδιαίτερα σημαντική μετά το 2013, προφανώς λόγω της σημαντικής αύξησης των τουριστικών ροών από τις χώρες του εξωτερικού προς την Ελλάδα (ΕΛΣΤΑΤ, 2018). Τέλος, ο συνολικός αριθμός των επιβατών εμφανίζει μια ανοδική τάση σε όλη τη χρονική περίοδο, ενώ η τάση αυτή είναι σχετικά μεγαλύτερη μετά το 2013.

Αξίζει να σημειωθεί ότι ενώ ο αριθμός των πτήσεων στο εσωτερικό της χώρας είναι περίπου ίσος με τον αριθμό των πτήσεων προς τις άλλες χώρες, ο αριθμός των μετακινούμενων εγχώριων επιβατών (Διάγραμμα 2) είναι σημαντικά πιο μικρός σε σχέση με τους επιβάτες των διεθνών πτήσεων. Αυτό δείχνει ότι τα χρησιμοποιούμενα αεροσκάφη στις εσωτερικές πτήσεις έχουν μικρότερη χωρητικότητα ως προς τα αεροσκάφη των διεθνών πτήσεων ή ακόμη η πληρότητα των αεροσκαφών στις εγχώριες πτήσεις είναι μικρότερη σε σχέση με την πληρότητα των αεροσκαφών στις διεθνείς πτήσεις.

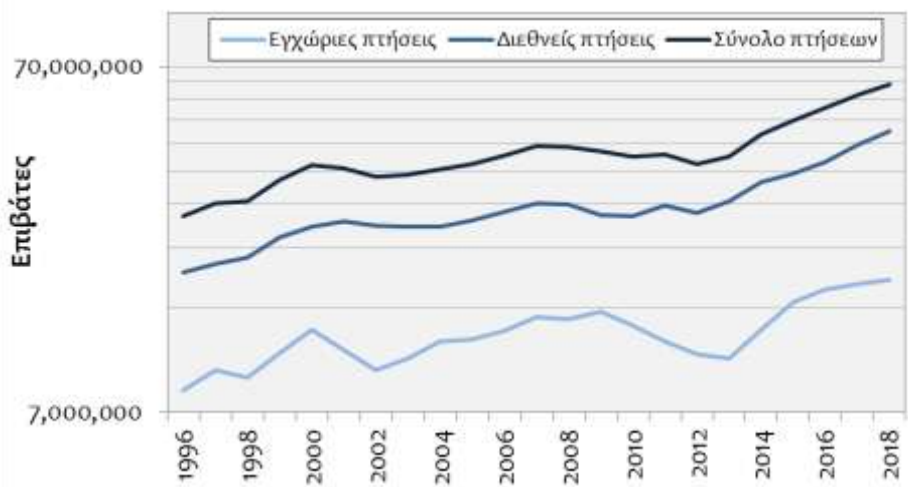

**Διάγραμμα 2:** Οι επιβάτες που μετακινήθηκαν στα αεροδρόμια της Ελλάδας την περίοδο 1996-2018 (οι τιμές στον κάθετο άξονα εμφανίζονται σε λογαριθμική κλίμακα $7x \cdot 10^6$, όπου $x=1,2,…,10$ είναι οι δευτερεύουσες γραμμές πλέγματος).

Τέλος, αναφορικά με τον αριθμό των εμπορευμάτων που μεταφέρθηκαν αεροπορικώς στις εγχώριες και διεθνείς πτήσεις κατά την περίοδο 1996-2018, αυτός, όπως απεικονίζεται στο Διάγραμμα 3, εμφανίζει μια συνεχή πτωτική πορεία. Η εξέλιξη αυτή δεν μπορεί να ερμηνευτεί εύκολα, δεδομένου ότι από το 1996 έως την έναρξη της οικονομικής κρίσης το 2009 η οικονομία της χώρας εμφάνισε αύξηση του ΑΕΠ. Συνεπώς, είναι αναμενόμενο, η αύξηση αυτή να συνοδεύεται με ανάλογη μεταβολή των εσωτερικών και εξωτερικών εμπορικών συναλλαγών. Μια μικρή ανοδική τάση στον αριθμό των μεταφερθέντων αεροπορικώς εμπορευμάτων από και προς το εξωτερικό εμφανίζεται μετά το 2013, πιθανόν λόγω της αύξησης των εξαγωγών της χώρας (ΕΛΣΤΑΤ, 2019).

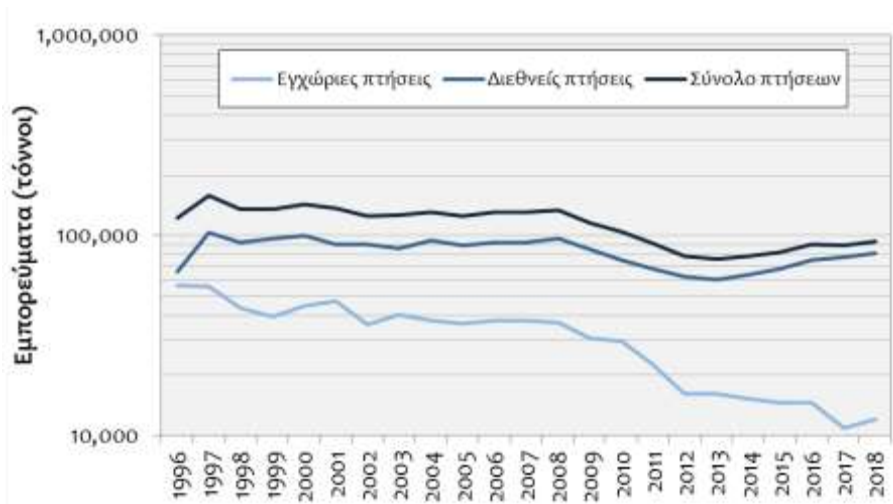

**Διάγραμμα 3:** Τα εμπορεύματα που μεταφέρθηκαν στα αεροδρόμια της Ελλάδας την περίοδο 1996-2018 (οι τιμές στον κάθετο άξονα εμφανίζονται σε λογαριθμική κλίμακα $\beta x$, όπου $\beta=10^4, 10^5$ είναι η αναγραφόμενη βάση και $x=1,2,\ldots,10$ οι δευτερεύουσες γραμμές πλέγματος).

Μια άλλη ερμηνεία που μπορεί να δοθεί στη συνεχή πτώση του μεταφορικού έργου εμπορευμάτων είναι ο ανταγωνισμός που δέχονται οι αεροπορικές μεταφορές από τα άλλα είδη μεταφορών (οδικές, ακτοπλοϊκές) (Tsiotas and Polyzos, 2015; 2018). Προφανώς το υψηλό μεταφορικό κόστος για μεγάλο όγκο ή βάρος εμπορευμάτων ευνοεί τους άλλους τύπους μεταφορών σε σχέση με τις αεροπορικές μεταφορές.

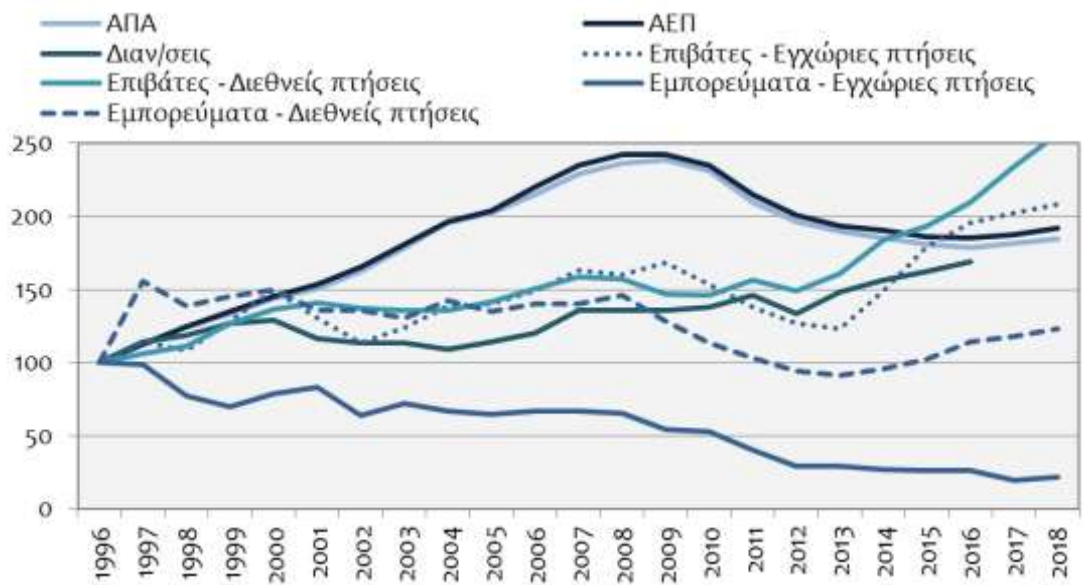

**Διάγραμμα 4:** Η εξέλιξη μεγεθών αεροπορικής δραστηριότητας και οικονομικών μεγεθών της Ελλάδας την περίοδο 1996-2018.

Τέλος, στο Διάγραμμα 4 εκτός από τη μεταβολή των μεταφερόμενων επιβατών και εμπορευμάτων στις πτήσεις εσωτερικού και εξωτερικού εμφανίζεται και η εξέλιξη του Ακαθάριστου Εγχώριου Προϊόντος (ΑΕΠ), της Ακαθάριστης Προστιθέμενης Αξίας (ΑΠΑ) (ΕΛΣΤΑΤ, 2019b) και του συνόλου των διανυκτερεύσεων στα ξενοδοχειακά καταλύματα κατά την περίοδο 1996 έως 2016 (ΕΛΣΤΑΤ, 2018). Είναι προφανές ότι τα παραπάνω μεγέθη αλληλοεπηρεάζονται και συνεπώς η εξέλιξη του ενός επιδρά στην εξέλιξη κάποιου από τα υπόλοιπα. Για τη διευκόλυνση στη σύγκριση, εξομοιώθηκαν τα εν λόγω μεγέθη που αντιστοιχούν στο έτος 1996 με 100 και στη συνέχεια υπολογίστηκαν οι σχετικές τιμές τους στα υπόλοιπα έτη. Παρατηρούμε ότι οι τιμές του ΑΕΠ και της ΑΠΑ εμφανίζουν ανοδικές τάσεις έως το 2009, έτος έναρξης της οικονομικής κρίσης, καθοδικές τάσεις έως το 2016 και ανοδικές τάσεις στη συνέχεια. Την τάση των δυο αυτών οικονομικών μεγεθών ακολουθούν τα μεγέθη που αντιστοιχούν στους επιβάτες και τα εμπορεύματα των εγχώριων και διεθνών πτήσεων αντίστοιχα. Τα μεγέθη που αντιστοιχούν στις διεθνείς πτήσεις επιβατών και τις διανυκτερεύσεις εμφανίζουν ανοδικές τάσεις σε όλη τη χρονική περίοδο, γεγονός που δείχνει την υψηλή εξάρτηση της διεθνούς επιβατικής αεροπορικής κίνησης με τον τουρισμό.

## 3. Η χωρική κατανομή και η μεταφορική δραστηριότητα των αεροδρομίων στην Ελλάδα

Η Ελλάδα διαθέτει εν λειτουργία 39 αεροδρόμια, εκ των οποίων τα αεροδρόμια των Αθηνών, της Θεσσαλονίκης, του Ηρακλείου, της Ρόδου και της Κέρκυρας ικανοποιούν περίπου το 85% της συνολικής αεροπορικής κίνησης της χώρας (ΕΛΣΤΑΤ, 2017). Από τα υπόλοιπα αεροδρόμια ένας σημαντικός αριθμός περιλαμβάνει κυρίως πτήσεις εξωτερικού. Ο αριθμός των αεροδρομίων που διαθέτει η χώρα θεωρείται αρκετά μεγάλος, αν συγκριθεί με τον αντίστοιχο αριθμό των άλλων ευρωπαϊκών χωρών (Tsiotas and Polyzos, 2015). Συγκεκριμένα, η Ελλάδα διαθέτει τον μεγαλύτερο αριθμό αερολιμένων ανά γεωγραφική επιφάνεια (3384 $Km^2$ ανά αερολιμένα), μετά τη Μάλτα και το Λουξεμβούργο που διαθέτουν ένα μόνο αεροδρόμιο, και είναι τέταρτη στην κατάταξη με τους λιγότερους κατοίκους ανά αεροδρόμιο μετά την Εσθονία, τη Φινλανδία και τη Σουηδία (Τσέκερης και Βογιατζόγλου, 2011). Ο κατακερματισμός του γεωγραφικού χώρου και οι γενικότερες ιδιομορφίες του έχουν ως αποτέλεσμα η χώρα να διαθέτει ένα μεγάλο αριθμό αεροδρομίων, όπως εμφανίζεται στο Χάρτη 1 (Tsiotas and Polyzos, 2015; Tsiotas et al., 2019a). Δηλαδή, οι θέσεις των αεροδρομίων εμφανίζουν μεγάλη χωρική διασπορά αν συσχετιστούν με τα σημεία των μεγάλων πληθυσμιακών της συγκεντρώσεων.

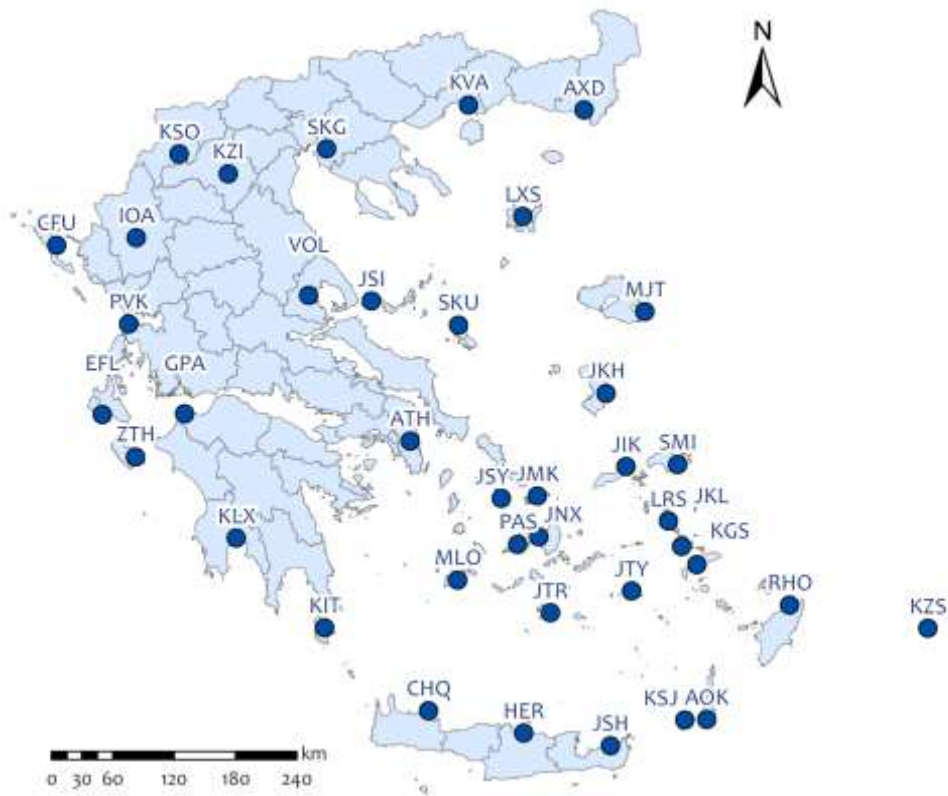

**Χάρτης 1:** Οι γεωγραφικές θέσεις των περιφερειακών αεροδρομίων στην Ελλάδα.

Τα πρώτα αεροδρόμια που κατασκευάστηκαν στην Ελλάδα είχαν ως σκοπό την ικανοποίηση κοινωνικών και οικονομικών αναγκών, αλλά χρησιμοποιήθηκαν και για στρατιωτικούς σκοπούς (Σκάγιαννης, 2008). Αυτό είχε ως αποτέλεσμα να χωροθετηθούν και να αναπτυχθούν αρχικά με κριτήρια την εξυπηρέτηση των στρατιωτικών τους σκοπών, ενώ στη συνέχεια μετατράπηκαν σε πολιτικά αεροδρόμια, χωρίς όμως να έχουν ικανοποιητικό σχεδιασμό που να στοχεύει στην αποτελεσματικότητα των αερομεταφορών και την ικανοποίηση των πραγματικών αναγκών. Η βασική σήμερα μεταφορική κίνηση των περιφερειακών αεροδρομίων στο μεγαλύτερο μέρος αφορά την εξυπηρέτηση των τουριστικών μετακινήσεων.

Η άποψη αυτή προκύπτει αβίαστα από μια μακροσκοπική θεώρηση του Χάρτη 1, όπου εμφανίζονται τα περιφερειακά αεροδρόμια της Ελλάδας, Σημειώνεται ότι από τον Χάρτη έχουν αφαιρεθεί τα δυο μεγάλα αεροδρόμια της χώρας, δηλαδή των Αθηνών και της Θεσσαλονίκης. Από το Χάρτη 1 διαπιστώνεται ότι εκτός τριών τα υπόλοιπα αεροδρόμια είναι χωροθετημένα σε παράκτιες περιοχές, γεγονός που έμμεσα δείχνει τη σύνδεσή τους με τον τουρισμό των περιοχών αυτών. Αναφορικά με τα δυο βασικά μεγέθη που δείχνουν την συνολική κίνηση κάθε αεροδρομίου, δηλαδή τον αριθμό των πτήσεων και τον αριθμό των μετακινηθέντων επιβατών, έχουν κατασκευαστεί τα Διαγράμματα 5 έως 8. Από τα διαγράμματα αυτά είναι εμφανές ότι τα αεροδρόμια μπορούν να διακριθούν σε δυο βασικές κατηγορίες, τα μεγάλα αεροδρόμια με περισσότερες από 10.000 πτήσεις ετησίως και τα μικρότερα αεροδρόμια με λιγότερες από 6.000 πτήσεις ετησίως.

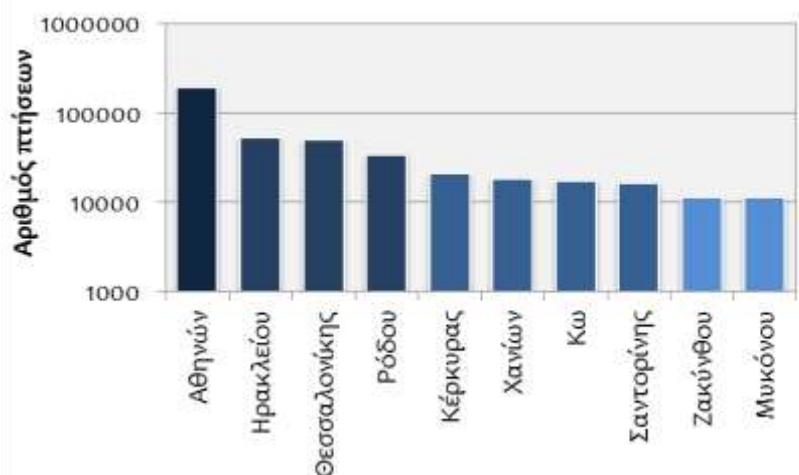

**Διάγραμμα 5:** Ο αριθμός πτήσεων για τα μεγαλύτερα αεροδρόμια της Ελλάδας το 2018.

Όπως προκύπτει από το Διάγραμμα 5, ο αριθμός των πτήσεων του αεροδρομίου των Αθηνών είναι 4-πλάσιος των πτήσεων των αμέσως επόμενων αεροδρομίων, δηλαδή του Ηρακλείου και της Θεσσαλονίκης. Ακολουθούν τα αεροδρόμια Ρόδου, Κερκύρας και Χανίων, των οποίων ο αριθμός πτήσεων είναι κατά πολύ μικρότερος και ίσος με το 10-15% των πτήσεων του αεροδρομίου των Αθηνών. Τη μεγαλύτερη συμμετοχή, ως προς το σύνολο των πτήσεων, κατέχει ο αερολιμένας Αθηνών με 39,60% και ακολουθούν τα αεροδρόμια Ηρακλείου και Θεσσαλονίκης με ποσοστά 10,77% και 10,24%, αντίστοιχα.

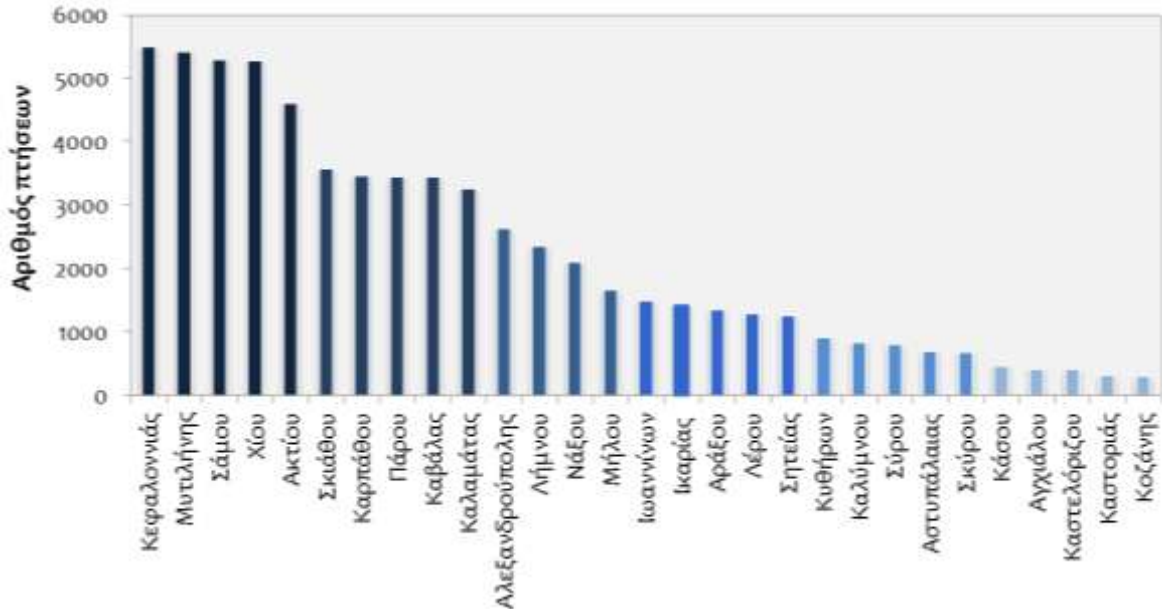

**Διάγραμμα 6:** Ο αριθμός πτήσεων για τα μικρότερα αεροδρόμια της Ελλάδας το 2018.

Τα υπόλοιπα περιφερειακά αεροδρόμια εμφανίζουν πολύ μικρότερο αριθμό πτήσεων, όπως απεικονίζεται στο Διάγραμμα 6. Χαρακτηριστικά αναφέρεται ότι τα 29 μικρότερα αεροδρόμια κατέχουν μόλις το 12.9 του συνόλου των πτήσεων των αεροδρομίων της χώρας, ενώ τα 5 μικρότερα αεροδρόμια κατέχουν μόλις το 0.3% του συνόλου των πτήσεων. Αναφορικά με τον αριθμό των επιβατών που μετακινούνται αεροπορικώς, από τα Διαγράμματα 7 και 8 προκύπτει ότι το αεροδρόμιο των Αθηνών καλύπτει το 38% ενώ τα αεροδρόμια Ηρακλείου, Θεσσαλονίκης και Ρόδου καλύπτουν συνολικά το 32% της συνολικής επιβατικής κίνησης. Διαφορετικά, τα 4 μεγαλύτερα αεροδρόμια της χώρας καλύπτουν το 70% της συνολικής επιβατικής κίνησης, ενώ τα υπόλοιπα 35 μόλις το 30% της επιβατικής κίνησης.

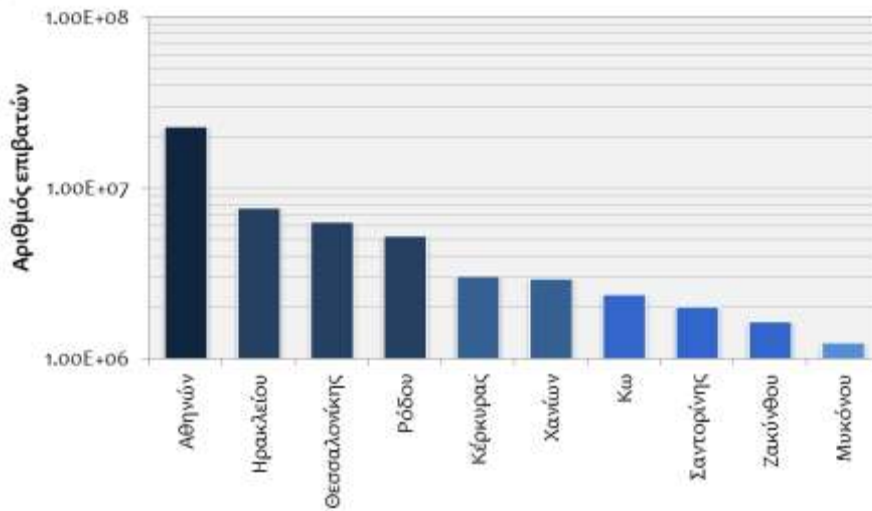

**Διάγραμμα 7:** Ο αριθμός επιβατών για τα μεγαλύτερα αεροδρόμια της Ελλάδας το 2018 (οι τιμές στον κάθετο άξονα εμφανίζονται σε λογαριθμική κλίμακα $βx$, όπου $β=10^6, 10^7$ είναι η αναγραφόμενη βάση και $x=1,2,…,10$ οι δευτερεύουσες γραμμές πλέγματος).

Από τη θεώρηση των Διαγραμμάτων 5 έως 8 προκύπτει μια κυριαρχία στις αεροπορικές μεταφορές ενός μικρού αριθμού αεροδρομίων, ενώ τα υπόλοιπα αεροδρόμια ικανοποιούν ένα πολύ μικρό ποσοστό της συνολικής μεταφορικής κίνησης. Ειδικότερα τα μικρά αεροδρόμια που βρίσκονται στη χερσαία χώρα (Αγχιάλου, Κοζάνης και Καστοριάς) ικανοποιούν ένα παρά πολύ μικρό μέρος της συνολικής ζήτησης, γεγονός που ίσως καθιστά με οικονομικούς όρους προβληματική ή ασύμφορη τη λειτουργία τους.

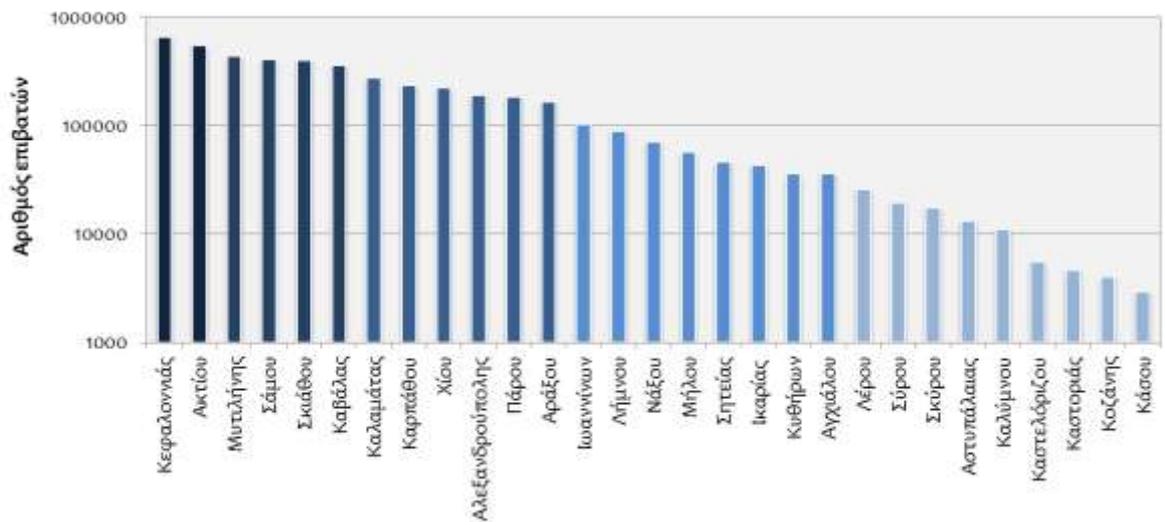

**Διάγραμμα 8:** Ο αριθμός επιβατών για τα μικρότερα αεροδρόμια της Ελλάδας το 2018

## 4. Ανάλυση της εποχικότητας της επιβατικής μεταφορικής δραστηριότητας των αεροδρομίων στην Ελλάδα

Η ζήτηση αεροπορικών υπηρεσιών, όπως και για πολλά άλλα αγαθά, εμφανίζει μια έντονη μηνιαία μεταβλητότητα εντός του έτους (Karlaftis, 2008; ΕΛΣΤΑΤ 2017). Η μεταβλητότητα αυτή χαρακτηρίζεται ως «εποχικότητα» και πολλούς συγγραφείς θεωρείται «λογική», αφού είναι συνδεδεμένη με τις τουριστικές ροές, οι οποίες είναι αυξημένες κατά τη θερινή περίοδο και μειωμένες τους υπόλοιπους μήνες του έτους (Kraft and Havlikova 2016). Όπως για πολλές επιχειρήσεις η εποχικότητα κατά κανόνα αποτελεί αρνητικό παράγοντα για τη λειτουργία και την οικονομική τους βιωσιμότητα, έτσι και για τη λειτουργία των αεροδρομίων η επίδρασή της είναι αρνητική. Η άνιση κατανομή της ζήτησης για αεροπορικές υπηρεσίες κατά τη διάρκεια του έτους δημιουργεί προβλήματα στη λειτουργία των αεροδρομίων με ανάλογες αρνητικές επιδράσεις στη βιωσιμότητά τους (Karlaftis, 2008; Kraft and Havlikova, 2016). Μια πρώτη αρνητική συνέπεια σχετίζεται με τα σταθερά έξοδα των αεροδρομίων, τα οποία κατά τη χειμερινή περίοδο λόγω της μικρής δραστηριότητας πιθανόν να μην καλύπτονται από τα συνολικά έσοδα. Για το λόγο αυτό σε αρκετές περιπτώσεις τα αεροδρόμια υπολειτουργούν κατά τη μεγαλύτερη περίοδο του έτους, με αρνητική επίδραση στην παραγωγικότητα και την αποδοτικότητά τους.

Η άλλη αρνητική συνέπεια αφορά στον προσδιορισμό του άριστου μεγέθους του αεροδρομίου. Είναι ευνόητο ότι, αν χρησιμοποιηθεί ως κριτήριο για τον προσδιορισμό του μεγέθους του αεροδρομίου η ικανοποίηση της ζήτησης αιχμής, τότε στις περιπτώσεις έντονης διακύμανσης της ζήτησης εντός του έτους η χρήση του κριτηρίου αυτού έχει ως αποτέλεσμα τη μη αξιοποίηση της υποδομής για μεγάλο χρονικό διάστημα. Στην αντίθετη περίπτωση, θα υπάρχουν χρονικές περίοδοι ή περίοδοι αιχμής στις οποίες δεν θα ικανοποιείται η συνολική ζήτηση για αεροπορικές μεταφορικές υπηρεσίες. Στις περιπτώσεις υπολειτουργίας ή κλεισίματος ενός αεροδρομίου για μια χρονική περίοδο, η αποδοτικότητα του επενδυμένου κεφαλαίου είναι πολύ μικρή ή μηδενική. Επίσης, δημιουργούνται κοινωνικά και οικονομικά προβλήματα με την απόλυση ή υποαπασχόληση του προσωπικού του, τα οποία επηρεάζουν αρνητικά τη συνολική λειτουργία του αεροδρομίου.

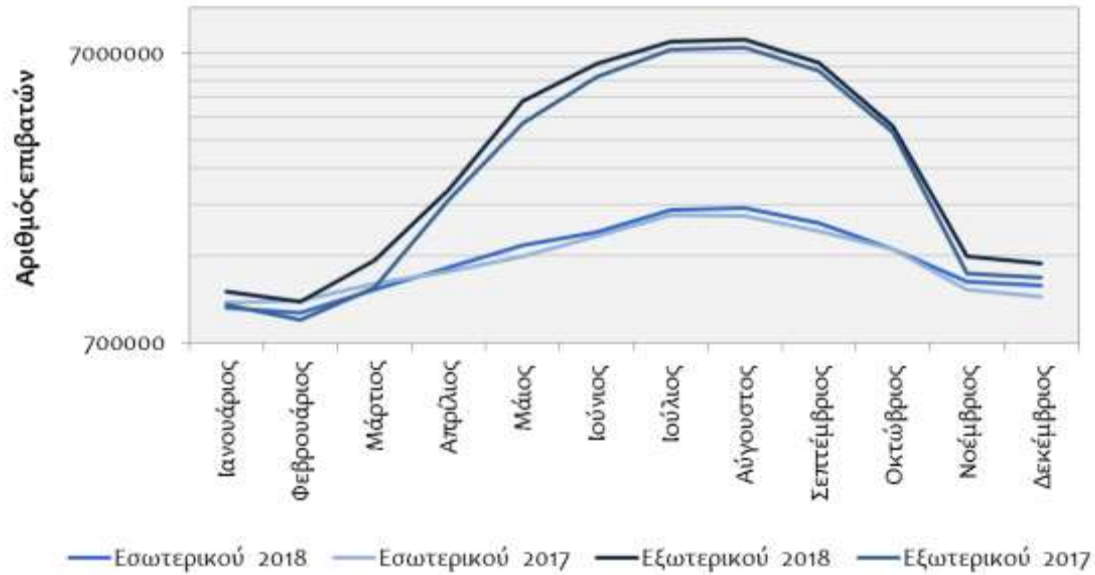

**Διάγραμμα 9:** Η κίνηση των επιβατών ανά μήνα στα αεροδρόμια της Ελλάδας την περίοδο 2017-2018

Για την απεικόνιση της μεταβλητότητας στη λειτουργία των αεροδρομίων της Ελλάδας έχει δημιουργηθεί το Διάγραμμα 9, όπου εμφανίζεται η αεροπορική κίνηση των ετών 2017 και 2018 (ΥΠΑ, 2019). Στο Διάγραμμα 9 διακρίνεται η έντονη εποχικότητά της αεροπορικής κίνησης στα αεροδρόμια της Ελλάδας. Συγκεκριμένα, παρατηρούμε τη συγκέντρωση της επιβατικής κίνησης του εξωτερικού στο 4-μηνο Ιουλίου-Σεπτεμβρίου, ενώ η επιβατική κίνηση των εσωτερικού δεν εμφανίζει έντονη εποχικότητα. Επίσης, στο Διάγραμμα 9 διακρίνουμε το μεγάλο ποσοστό που καλύπτουν στη συνολική αεροπορική κίνηση οι πτήσεις του εξωτερικού. Τόσο το μεγάλο ποσοστό της επιβατικής κίνησης του εξωτερικού όσο και η εποχικότητά της μας οδηγούν στο συμπέρασμα ότι η αεροπορική κίνηση στην Ελλάδα είναι πολύ στενά συνδεδεμένη με την τουριστική κίνηση.

Θα επιδιωχθεί στη συνέχεια η ποσοτική έκφραση της εποχικότητας της επιβατικής κίνησης των αεροδρομίων της Ελλάδας και η σύγκριση των τιμών της. Έχοντας ως βάση το Διάγραμμα 8 και τη μορφή που έχει η καμπύλη μηνιαίας κατανομής της επιβατικής κίνησης, θα υπολογιστεί ο βαθμός συγκέντρωσης των τιμών γύρω από τη μέση τιμή. Ένα μέτρο για τον έλεγχο του βαθμού της συγκέντρωσης των τιμών γύρω από τη μέση τιμή είναι ο συντελεστής κύρτωσης α, ο οποίος υπολογίζεται από τη σχέση:

$$a = \frac{\frac{1}{n}\sum_{i=1}^{n}(x_i - \bar{x})^4}{\{\sqrt{\frac{1}{n}\sum_{i=1}^{n}(x_i - \bar{x})^2}\}^4} \qquad (1),$$

όπου:
n = Ο αριθμός των μηνιαίων παρατηρήσεων.
$x_i$ = Η τιμή της παρατήρησης i.
$\bar{x}$ = Η μέση τιμή των παρατηρήσεων i=1÷n.

Επειδή για τις κανονικές κατανομές έχουμε α=3, η μέτρηση της κυρτότητας πραγματοποιείται με χρήση της διαφοράς α-3, η οποία για λεπτόκυρτες κατανομές παίρνει θετικές τιμές, ενώ για πλατύκυρτες γίνει αρνητική (Montgomery and Runger, 2003). Επιπλέον, θα υπολογιστεί ο συντελεστής Gini, ο οποίος αποτελεί έναν διαφορετικό τρόπο μέτρησης της ανισοκατανομής. Ο υπολογισμός του συντελεστή Gini προκύπτει από την καμπύλη συγκέντρωσης του Lorenz, η οποία χρησιμοποιείται για τη διαγραμματική απεικόνιση των χωρικών ανισοτήτων στην κατανομή ορισμένων μεγεθών εντός μιας περιφέρειας ή μιας χώρας (Πολύζος 2019a). Βασίζεται στον υπολογισμό της σχέσης της ποσοστιαίας αθροιστικής κατανομής ενός μεγέθους με την κατανομή των μονάδων που αντιστοιχούν σε αυτή. Δυο χαρακτηριστικές απεικονίσεις που αφορούν στην μηνιαία κατανομή επιβατών σε μια περιοχή με τη χρήση της καμπύλης Lorenz εμφανίζονται στο Διάγραμμα 10.

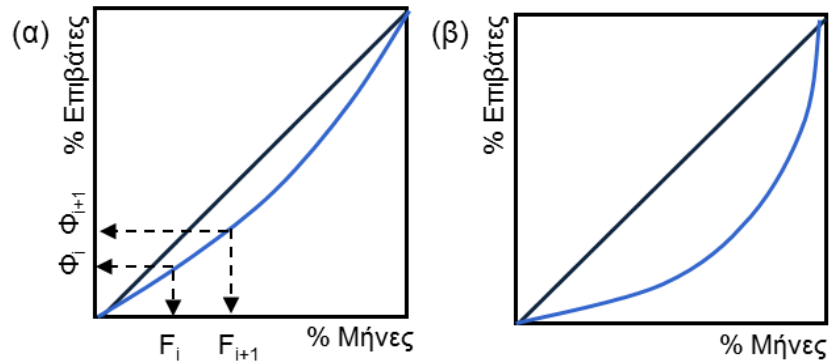

**Διάγραμμα 10:** Η καμπύλη Lorenz του διαγράμματος (α) αντιστοιχεί σε μικρές ανισότητες, ενώ του διαγράμματος (β) αντιστοιχεί σε μεγάλες ανισότητες.

Με βάση το Διάγραμμα 9(α), ο συντελεστής Gini μπορεί να υπολογιστεί από τη σχέση (Πολύζος, 2019b; Tsiotas et al., 2019b):

$$G_i = 1 - \sum_{i=0}^{n-1}(F_{i+1} - F_i)(\Phi_{i+1} - \Phi_i) \quad (2).$$

Χρησιμοποιώντας τις σχέσεις (1) και (2) υπολογίζουμε τους συντελεστές κύρτωσης α και Gini ($G_i$) για τα αεροδρόμια της Ελλάδας που αφορούν την μηνιαία κατανομή των επιβατών για τις πτήσεις του εσωτερικού και εξωτερικού. Τα αποτελέσματα των υπολογισμών εμφανίζονται στον Πίνακα 1. Από τα αποτελέσματα του Πίνακα 1 προκύπτει ότι υπάρχουν αεροδρόμια με μεγάλη εποχικότητα στις πτήσεις εσωτερικού ή τις διεθνείς πτήσεις και ορισμένα με μικρή εποχικότητα. Παρατηρούμε ότι ένας σημαντικός αριθμός αεροδρομίων έχουν συντελεστή κύρτωσης α>3 και συνεπώς αντιστοιχούν σε λεπτόκυρτες κατανομές, καθώς και υψηλές τιμές του συντελεστή Gini. Αντίθετα πολύ λιγότερα αντιστοιχούν σε πλατύκυρτες κατανομές και έχουν χαμηλές τιμές του συντελεστή κύρτωσης και συντελεστή Gini. Οι τιμές αυτές των συντελεστών δείχνουν ποια αεροδρόμια χαρακτηρίζονται από το μειονέκτημα της εποχικότητας και συνεπώς απαιτούνται γενικότερες δράσεις για τη μείωσή της.

**Πίνακας 1**
Έλεγχος του βαθμού εποχικότητας της κίνησης των αεροδρομίων της Ελλάδας με χρήση των συντελεστών κύρτωσης και Gini

| Αεροδρόμιο | Συντελεστής κύρτωσης α | | Συντελεστής Gini $G_i$ | |
|---|---|---|---|---|
| | Πτήσεις εσωτερικού | Διεθνείς πτήσεις | Πτήσεις εσωτερικού | Διεθνείς πτήσεις |
| Αράξου | 3,555 | 1,508 | 0,793 | 0,612 |
| Ακτίου | 2,828 | 1,640 | 0,649 | 0,607 |
| Αλεξανδρούπολης | 1,863 | 3,226 | 0,118 | 0,718 |
| Αστυπάλαιας | 2,573 | - | 0,460 | - |
| Αγχιάλου | 4,123 | 1,845 | 0,818 | 0,582 |
| Ζακύνθου | 3,054 | 1,545 | 0,260 | 0,603 |
| Ηρακλείου | 1,771 | 1,352 | 0,095 | 0,512 |
| Θεσσαλονίκης | 2,461 | 1,557 | 0,050 | 0,246 |
| Ικαρίας | 2,861 | - | 0,306 | - |
| Ιωαννίνων | 1,924 | 1,609 | 0,084 | 0,544 |
| Καβάλας | 1,789 | 1,710 | 0,062 | 0,424 |
| Καλαμάτας | 1,943 | 1,431 | 0,261 | 0,531 |
| Καλύμνου | 2,855 | - | 0,294 | - |
| Καρπάθου | 3,259 | 1,813 | 0,314 | 0,643 |
| Κάσου | 2,793 | - | 0,394 | - |
| Καστελόριζου | 2,207 | - | 0,473 | - |
| Καστοριάς | 2,590 | 10,093 | 0,089 | 0,917 |
| Κέρκυρας | 2,412 | 1,499 | 0,159 | 0,568 |
| Κεφαλονιάς | 2,709 | 1,505 | 0,422 | 0,592 |
| Κοζάνης | 1,630 | - | 0,071 | - |
| Κυθήρων | 2,244 | 1,412 | 0,384 | 0,600 |
| Κω | 2,685 | 1,322 | 0,189 | 0,559 |
| Λέρου | 2,094 | - | 0,371 | - |
| Λήμνου | 3,337 | 1,518 | 0,251 | 0,630 |
| Μήλου | 1,548 | - | 0,398 | - |
| Μυκόνου | 1,482 | 2,039 | 0,458 | 0,618 |
| Μυτιλήνης | 2,381 | 1,589 | 0,110 | 0,619 |
| Νάξου | 1,673 | - | 0,405 | - |
| Πάρου | 2,175 | 2,940 | 0,438 | 0,713 |
| Ρόδου | 2,306 | 1,378 | 0,092 | 0,551 |
| Σάμου | 3,222 | 1,573 | 0,133 | 0,612 |
| Σαντορίνης | 1,243 | 1,768 | 0,296 | 0,558 |
| Σητείας | 2,910 | 1,579 | 0,243 | 0,596 |
| Σκιάθου | 3,173 | 1,864 | 0,454 | 0,654 |
| Σκύρου | 2,472 | 3,667 | 0,428 | 0,779 |
| Σύρου | 2,746 | - | 0,265 | - |
| Χανίων | 2,381 | 1,337 | 0,094 | 0,518 |
| Χίου | 2,845 | 1,976 | 0,127 | 0,691 |
| Αθηνών | 1,659 | 1,700 | 0,160 | 0,162 |

Όπως προαναφέρθηκε, η εποχικότητα αποτελεί πρόβλημα για την αποτελεσματική λειτουργία και τη βιωσιμότητα των αεροδρομίων της Ελλάδας. Για τη μείωση της εποχικότητας και την πιο ομαλή κατανομή της αεροπορικής κίνησης και της δραστηριότητας των αεροδρομίων στη διάρκεια του έτους θα πρέπει να υπάρξουν δράσεις που θα αφορούν κυρίως στην αύξηση της τουριστικής δραστηριότητας εκτός της θερινής περιόδου και της γενικότερης οικονομικής ανάπτυξης των νησιωτικών περιφερειών της χώρας. Αναφορικά με την εποχικότητα του τουρισμού στην Ελλάδα, η οποία επηρεάζει καθοριστικά την κατανομή της αεροπορικής κίνησης εντός του έτους, αυτή οφείλεται σε δύο γενικές αιτίες. Η πρώτη αφορά στις γενικότερες κλιματολογικές συνθήκες που επικρατούν στη χώρα, οι οποίες ευνοούν τον τουρισμό των διακοπών ή μαζικό τουρισμό και γενικότερα τη διάρθρωση του τουρισμού που επικρατεί στην Ελλάδα.

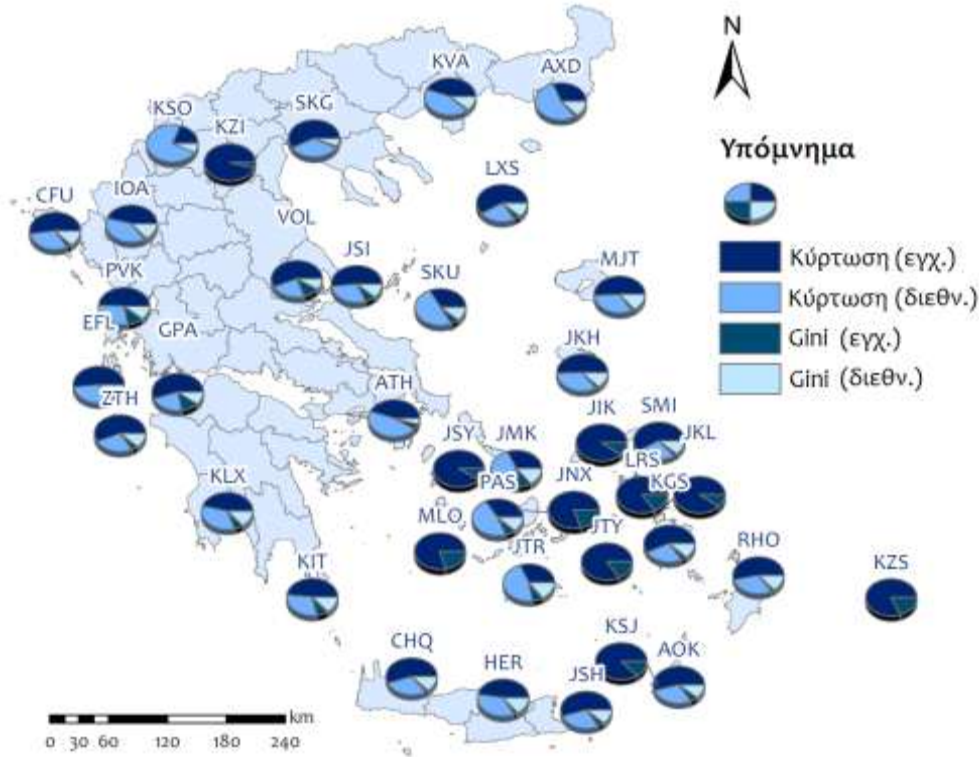

**Χάρτης 2:** Γεωγραφική κατανομή του βαθμού εποχικότητας της κίνησης των αεροδρομίων της Ελλάδας με χρήση των συντελεστών κύρτωσης και Gini.

Γενικά, σε όλες τις μεσογειακές χώρες οι υψηλές θερμοκρασίες των θερινών μηνών ευνοούν την αύξηση του τουρισμού των ακτών και την όξυνση του φαινομένου της εποχικότητας. Η δεύτερη αιτία συνδέεται με τη θεσμική οργάνωση και ειδικότερα την οργάνωση της σχολικής ζωής και την τοποθέτηση των διακοπών στη θερινή περίοδο, το εορτολόγιο και τη θέση των θρησκευτικών εκδηλώσεων στο ημερολόγιο, τα ήθη και τα έθιμα που επικρατούν στη χώρα. Υπό μια γενική θεώρηση, η θεσμική οργάνωση είναι άμεσα συνδεδεμένη με τις κλιματολογικές συνθήκες, αφού η οργάνωση πολλών εκδηλώσεων ευνοείται από τις καλές κλιματολογικές συνθήκες που επικρατούν τη θερινή περίοδο. Επίσης, η ανάπτυξη των νησιωτικών περιφερειών και η τόνωση της οικονομικής τους δραστηριότητας, θα ενισχύσει τόσο τις εσωτερικές όσο και τις διεθνείς αεροπορικές μετακινήσεις. Το «περιφερειακό πρόβλημα» είναι ιδιαίτερα έντονο στην Ελλάδα, με αποτέλεσμα οι μεγάλες συγκεντρώσεις πληθυσμού και οικονομικών δραστηριοτήτων στα δυο μεγάλα αστικά κέντρα της Αθήνας και της Θεσσαλονίκης να μην επιτρέπουν την ανάπτυξη των περιφερειών και ειδικότερα των γεωγραφικά απομονωμένων περιοχών, όπως είναι οι νησιωτικές περιοχές.

## 5. Η σημασία των περιφερειακών αεροδρομίων για την τοπική και περιφερειακή ανάπτυξη

Οι εξελίξεις στη διεθνή οικονομία και η αυξανόμενη παγκοσμιοποίηση, όπως αυτή έμμεσα προσδιορίζεται από τις μεταβολές που επιφέρουν οι μετακινήσεις πληθυσμού και οι υλικές ή άυλες «ροές» εμπορευμάτων, τεχνολογίας, πολιτισμού και γνώσης, καθιστούν τα αεροδρόμια κατηγορία μεταφορικών υποδομών με συνεχώς αυξανόμενη σημασία (Andrew and Bailey, 1996; Green, 2007; Mukkala and Tervo, 2013). Γενικότερα, οι αεροπορικές μεταφορές και οι υποδομές τους αποτελούν σημαντικό παράγοντα βελτίωσης του οικονομικού δυναμικού μιας περιοχής και των προοπτικών της οικονομικής της ανάπτυξης (Alkaabi and Debbage, 2007; Zak and Getzner, 2014; Πολύζος, 2019b). Από τη δεκαετία του '70 και μετά η ανάπτυξη των αερομεταφορών είναι σημαντική και καλύπτει σχεδόν όλες τις χώρες του κόσμου, παρέχοντας τη δυνατότητα μεταφοράς σε μέρη όπου άλλα μεταφορικά μέσα δεν μπορούν να φτάσουν (Kraft and Havlikova 2016). Γενικά, οι αερομεταφορές αποτελούν ανταγωνιστικό μέσο για μεταφορές επιβατών σε μεγάλες αποστάσεις και για εμπορεύματα μικρού βάρους και όγκου. Το κόστος αερομεταφοράς συνεχώς μειώνεται, με αποτέλεσμα να αυξάνεται η εμβέλεια διάθεσης των παραγόμενων προϊόντων πολλών περιοχών ταυτόχρονα με τη σημασία των αεροπορικών υποδομών για την τοπική ανάπτυξη.

Η επίδραση των αεροδρομίων στην τοπική ανάπτυξη και ο ρόλος τους στην περιφερειακή ανάπτυξη έχει διερευνηθεί εμπειρικά σε πολλές έρευνες διεθνώς. Παρά το γεγονός ότι υπάρχει μια ποικιλία απόψεων για το εύρος και το μέγεθος της εν λόγω επίδρασης, τα τελικά συμπεράσματα στα οποία καταλήγει η πλειονότητα των ερευνών συγκλίνουν στην άποψη ότι η λειτουργία ενός αεροδρομίου βελτιώνει την παραγωγικότητα πολλών οικονομικών δραστηριοτήτων μιας περιφέρειας και αυξάνει την ελκυστικότητά της για εγκατάσταση παραγωγικών δραστηριοτήτων (Andrew and Bailey, 1996; Green, 2007). Όμως, ενώ υπάρχει συνήθως ισχυρή συσχέτιση μεταξύ μεταφορικών υποδομών και της οικονομικής ανάπτυξης, η μεταξύ τους αιτιώδης σχέση δεν είναι πάντοτε απολύτως σαφής και διακριτή (Πολύζος, 2019b). Για τα αεροδρόμια και τις αεροπορικές υποδομές, παρά την ύπαρξη εμφανούς σχέσης τους με την οικονομική ανάπτυξη, η αδυναμία υπολογισμού του βαθμού επίδρασης του αιτίου στο αιτιατό καθιστά ιδιαίτερα δύσκολη την ex ante μελέτη των οικονομικών επιδράσεων των αεροδρομίων στην ανάπτυξη των περιοχών (Green, 2007).

Οι περισσότερες έρευνες υποστηρίζουν ότι, στην περίπτωση των απομακρυσμένων περιοχών, οι αερομεταφορές παίζουν καίριο ρόλο στην ανάπτυξή τους. Βασική κατάληξη των προσεγγίσεων είναι ότι, το είδος και η κλίμακα των οικονομικών επιπτώσεων των αεροδρομίων, εξαρτάται από τα ιδιαίτερα χαρακτηριστικά των περιφερειών που εξυπηρετούν και το ρόλο τους στην ευρύτερη οικονομία (Button and Lall, 1999; Mukkala and Tervo, 2013; Πολύζος, 2019b). Γενικά, όπως για όλες τις μεταφορικές υποδομές, έτσι και για τις αεροπορικές ισχύει ο γενικός κανόνας, πως «..*είναι αναγκαίες, αλλά όχι ικανές προϋποθέσεις για την οικονομική ανάπτυξη μιας περιοχής..*» (Green, 2007; Πολύζος, 2019b). Σε αρκετές περιπτώσεις έχει παρατηρηθεί μια «κυκλική» λειτουργία της ανάπτυξης και των μεταφορικών υποδομών. Έτσι, η οικονομική ανάπτυξη μπορεί να ωθήσει μια περιοχή στη βελτίωση των μεταφορικών της υποδομών. Στη συνέχεια οι νέες υποδομές επιταχύνουν τους ρυθμούς ανάπτυξης, αυξάνουν την προσβασιμότητα της περιοχής, ενισχύουν την ικανότητα για έλξη νέων επενδύσεων και την ανταγωνιστικότητά της. Οι εξελίξεις αυτές δημιουργούν την ανάγκη για περαιτέρω βελτίωση των μεταφορικών της υποδομών και συνεχίζεται ο «κύκλος» βελτίωσης υποδομών και ανάπτυξης (Mukkala and Tervo, 2013; Πολύζος, 2015; Πολύζος, 2019a,b). Σε πολλές μελέτες κατέληξαν στο συμπέρασμα ότι, η διακίνηση ενός εκατομμυρίου επιβατών σε ένα αεροδρόμιο μεσαίου και μεγάλου μεγέθους, επιφέρει την άμεση δημιουργία 1000 θέσεων εργασίας (Graham, 2008), ενώ άλλοι μελετητές καταλήγουν σε πιο μέτριους υπολογισμούς (Zak και Getzner, 2014).

Σε μια γενική θεώρηση, η συνολική επίδραση των αεροδρομίων στην τοπική ανάπτυξη εξαρτάται από ορισμένους παράγοντες και κάποια γενικά ή ειδικά χαρακτηριστικά των περιφερειών, τα βασικότερα εκ των οποίων αφορούν τα *οικονομικά, κοινωνικά και γεωγραφικά χαρακτηριστικά* των *περιοχών που εξυπηρετούνται* από το αεροδρόμιο και την *εφαρμοζόμενη κρατική πολιτική*. Ειδικότερα, *τα χαρακτηριστικά του αεροδρομίου* και συγκεκριμένα η ικανότητά του να παρέχει όλες τις αεροπορικές υπηρεσίες, καθώς και η ανταγωνιστικότητα ή η συμπληρωματικότητά του με μεταφορικές ή άλλες υποδομές της περιοχής.

Στα *οικονομικά χαρακτηριστικά* μπορούν να ενταχθούν η παραγωγική ειδίκευση της οικονομίας των περιοχών, η εξάρτηση της οικονομίας τους από το μεταφορικό κόστος, οι πλουτοπαραγωγικοί πόροι που διαθέτει κάθε περιοχή και οι οποίοι δυνητικά μπορούν να αξιοποιηθούν, καθώς και άλλα ειδικά χαρακτηριστικά, όπως είναι η παραγωγικότητα και η ανταγωνιστικότητα της οικονομίας, η συμπληρωματικότητα της οικονομίας κάθε περιοχής με τις οικονομίες που συναλλάσσεται εμπορικά, η κινητικότητα των συντελεστών παραγωγής κ.λπ. Για παράδειγμα, η υψηλή εξάρτηση της οικονομίας και κυρίως του τουρισμού των ελληνικών νησιών από τις αεροπορικές μεταφορές, καθιστούν την ύπαρξη αεροδρομίου σε κάθε νησιωτική περιοχή αναγκαία προϋπόθεση για την οικονομική της ανάπτυξη.

Στα *κοινωνικά χαρακτηριστικά* μπορούν να ενταχθούν το κοινωνικό κεφάλαιο κάθε περιοχής, τα δημογραφικά χαρακτηριστικά του, η ηλικιακή πυραμίδα, η επιχειρηματική παράδοση, η τάση για ανάληψη επιχειρηματικού ρίσκου, κ.λ.π. Δεν θα πρέπει να αγνοηθεί το γεγονός ότι τα αεροδρόμια έμμεσα παίζουν καταλυτικό ρόλο σε πολλές περιοχές για τις κοινωνικές αλλαγές, αφού ευνοούν την περιφερειακή κοινωνική συνοχή και τη βελτίωση της ελκυστικότητας πολλών περιοχών για παραμονή ή εγκατάσταση κατοίκων υψηλού κοινωνικού ή μορφωτικού επιπέδου (Button and Lall, 1999; Zak and Getzner, 2014).

Στα *γεωγραφικά χαρακτηριστικά* μπορούν να ενταχθούν η γεωγραφική θέση κάθε περιφέρειας, ο βαθμός απομόνωσης ή διασύνδεσης με άλλες περιοχές, η μορφολογία του εδάφους κ.λπ. Για πολλές απομονωμένες γεωγραφικά περιοχές, τα γεωγραφικά μειονεκτήματα μπορούν να μετριαστούν σημαντικά με την ανάπτυξη των αεροπορικών μεταφορών, αφού έτσι αποδυναμώνονται οι αρνητικές επιδράσεις των μεγάλων αποστάσεων και της απομόνωσης.

Η *κρατική πολιτική* που εφαρμόζεται και σχετίζεται με τις κρατικές παρεμβάσεις που μπορούν να αναδείξουν τα πλεονεκτήματα της περιφέρειας και να ενισχύσουν τη θετική συμβολή των αεροπορικών υποδομών στην τοπική ανάπτυξη ή το αντίθετο. Ως τέτοιες παρεμβάσεις μπορούν να αναφερθούν η κατασκευή συμπληρωματικών ή ανταγωνιστικών υποδομών, η γενικότερη αναπτυξιακή και περιφερειακή πολιτική που επηρεάζει τη θέση κάθε περιοχής στο χωρικό ανταγωνισμό δημιουργώντας συγκριτικά πλεονεκτήματα, τα αναπτυξιακά κίνητρα (επιδοτήσεις, φοροαπαλλαγές), η γενικότερη φορολογική πολιτική κ.λπ. Γενικά, για κάθε περιοχή που εμφανίζει οικονομική ανάπτυξη, ενώ η οικονομία της είναι «εξωστρεφής», δηλαδή εξαρτάται σε μεγάλο βαθμό από τις εισαγωγές και τις εξαγωγές, αυξάνεται ανάλογα και η ζήτηση αεροπορικών υπηρεσιών. Θεωρείται, επίσης, ότι η προσβασιμότητα στις αεροπορικές μεταφορές είναι μία από τις πολλές προϋποθέσεις για την αύξηση της ανάπτυξης και της ανταγωνιστικότητας μιας περιοχής. Οι αεροπορικές υπηρεσίες παρέχουν έναν έγκαιρο και αξιόπιστο μηχανισμό μεταφοράς ατόμων, αγαθών και υπηρεσιών από τον ένα τόπο στον άλλο σε ένα παγκοσμιοποιημένο περιβάλλον.

Παρά το γεγονός ότι για πολλούς τα αεροδρόμια μπορεί να φαίνονται οργανικά αποσυνδεδεμένα από τις περιοχές ή τις πόλεις στις οποίες βρίσκονται, στην πραγματικότητα αποτελούν κρίσιμη συνιστώσα της σύνδεσης των ανθρώπων και των τόπων και ως εκ τούτου συμβάλλουν σημαντικά στην περιφερειακή οικονομική ανάπτυξη (Florida et al., 2015). Για πολλούς, τα αεροδρόμια είναι κάτι περισσότερο από μια απλή μεταφορική υποδομή. Συμβάλλουν στη βελτίωση της εικόνας μιας πόλης ή μιας περιοχής, αφού παρέχουν «ανώτερη πρόσβαση (superior access) στις παγκόσμιες ροές ανθρώπων, αγαθών, χρημάτων και πληροφοριών» (O'Connor, 2003).

Η συνεχής μεγέθυνση των αεροδρομίων και η στέγαση εκτός αυτών διαφόρων υπηρεσιών μεταφοράς και άλλων συμπληρωματικών υπηρεσιών, οδήγησαν στην εμφάνιση μιας νέας τουριστικής τάσης που ονομάστηκε "terminal tourism" (τουρισμός τερματικών σταθμών). Η τάση αυτή πρωτοεμφανίστηκε στις ΗΠΑ και αφορά την προώθηση του τουρισμού των αεροδρομίων και περιλαμβάνει την περιήγηση συνοδών των επιβατών ή άλλων επισκεπτών στους τερματικούς σταθμούς. Το είδος αυτών των επισκεπτών διέρχεται από τους ελέγχους ασφαλείας και κατά την παραμονή του στο αεροδρόμιο έχει τη δυνατότητα να επισκεφτεί εκθέσεις, να κάνει αγορές, να γευματίσει, να συμμετάσχει σε άλλες δραστηριότητες κ.λπ. Στην Ελλάδα, η συμβολή του αεροδρομίου στην ανάπτυξη της περιοχής όπου βρίσκεται δια μέσου της έλξης και άλλων συμπληρωματικών οικονομικών δραστηριοτήτων, απεικονίζεται πολύ καθαρά στην εξέλιξη της περιοχής των Σπάτων μετά την κατασκευή του αεροδρομίου «Ελ. Βενιζέλος».

## 6. Συμπεράσματα και προτάσεις πολιτικής

Σύμφωνα με όλες τις εκτιμήσεις, η ανάπτυξη των διαπεριφερειακών συγκοινωνιακών υποδομών αποτελεί τη βάση για την οικονομική ανάπτυξη κάθε περιφέρειας. Ειδικότερα για τις πιο απομονωμένες γεωγραφικά περιοχές, οι αεροπορικές υποδομές αποτελούν έργα ζωτικής σημασίας για την ανάπτυξή τους, αφού βοηθούν στην υπέρβαση των εμποδίων που επιφέρει η γεωγραφική απομόνωση. Για πολλές περιοχές η σχέση μεταξύ των μεταφορικών υποδομών τους και της οικονομικής ανάπτυξης υπερβαίνει το στενό πλαίσιο που έμμεσα ορίζεται από το βασικό σκοπό των εν λόγω υποδομών και αφορά στη μεταφορά ανθρώπων και αγαθών από τον ένα τόπο στον άλλο.

Για τις νησιωτικές περιοχές της Ελλάδας, οι οποίες κατέχουν και τον μεγαλύτερο αριθμό αεροδρομίων της χώρας, οι αεροπορικές μεταφορές αποτελούν «κλειδί» για την οικονομική τους επιβίωση, αφού οι οικονομικές τους δραστηριότητες συνδέονται στενά με αυτές. Η ανάλυση που προηγήθηκε έδειξε σαφώς την υψηλή εξάρτηση του τουρισμού της χώρας, που αποτελεί τη σημαντικότερη οικονομική δραστηριότητα των νησιωτικών περιοχών, από τα αεροδρόμια και τις αεροπορικές μεταφορές. Συνεπώς, η διάθεση ικανοποιητικών αεροπορικών υποδομών ενισχύει τα συγκριτικά πλεονεκτήματα των τουριστικών περιοχών της χώρας έναντι των ανταγωνιστών της και βοηθά σημαντικά στη διατήρηση και την αύξηση των τουριστικών ροών. Επιπλέον, οι αεροπορικές υποδομές διασφαλίζουν τη σύνδεση των νησιωτικών περιοχών με τις ηπειρωτικές περιφέρειες και βοηθούν στην άρση της γεωγραφικής τους απομόνωσης.

Η γεωπολιτική θέση της Ελλάδας επιβάλει την περαιτέρω ανάπτυξη των αεροδρομίων και του αερομεταφορικού της δικτύου, ώστε να εξασφαλιστεί η απαιτούμενη χωρητικότητα για την εξυπηρέτηση της εκτιμώμενης μελλοντικής ζήτησης. Σύμφωνα με τις προβλέψεις και ύστερα από την εφαρμογή των κατάλληλων αναπτυξιακών στρατηγικών η μελλοντική ζήτηση για αερομεταφορές προβλέπεται να είναι αυξανόμενη. Επιπλέον, η προβολή της τάσης που εμφανίζεται στα διαγράμματα που παραπάνω παρουσιάστηκαν δείχνει αυξητικές προοπτικές στην κίνηση των αεροπορικών μεταφορών.

Ο εκσυγχρονισμός των υποδομών των αεροδρομίων της χώρας και οι αναβαθμίσεις των παρεχόμενων υπηρεσιών θα βοηθήσουν στην άνοδο του επιπέδου ικανοποίησης του επιβατικού κοινού και στην μεγαλύτερη αποδοτικότητα των αεροδρομίων. Αυτό θα δώσει τη δυνατότητα για διεύρυνση των συνδέσεων όλων των αεροδρομίων με τα κύρια πολιτικά και οικονομικά κέντρα της Ελλάδας και του εξωτερικού και την αύξηση της συμβολής τους στην τοπική ανάπτυξη. Τέλος, αξίζει το ενδιαφέρον να διερευνηθεί η σκοπιμότητα εφαρμογής ενός Hub-and-Spoke (Tsiotas, 2019) συστήματος στο αεροπορικό δίκτυο για κάποια αεροδρόμια της χώρας.

## Βιβλιογραφία / References